\documentclass[twocolumn]{aastex62}

\usepackage{amsfonts}
\usepackage{amsmath}
\usepackage{amssymb}
\usepackage{bm}% bold math
\usepackage{color}
\usepackage{dcolumn}% Align table columns on decimal point
\usepackage{epsfig}
\usepackage{eucal}
\usepackage{float}
\usepackage{graphicx}% Include figure files
\usepackage[latin1]{inputenc}
\usepackage{microtype}
\usepackage{subfigure}
\usepackage{tikz}
\usepackage{ulem}%delete line package
\usepackage{verbatim}
\usepackage{xcolor}
\usepackage{cancel}

\shortauthors{Cai et al.}

\begin{document}

%\title{Model-independent reconstruction of $f(T)$ gravity from Gaussian Processes and alleviation of the $H_0$ tension}
\title{Model-independent reconstruction of $f(T)$ gravity from Gaussian Processes}

\author{Yi-Fu Cai}
\affiliation{Department of Astronomy, School of Physical Sciences, University of Science and Technology of China, Hefei, Anhui 230026, China}
\affiliation{CAS Key Laboratory for Research in Galaxies and Cosmology, University of Science and Technology of China, Hefei, Anhui 230026, China}
\affiliation{School of Astronomy and Space Science, University of Science and Technology of China, Hefei, Anhui 230026, China}

\author{Martiros Khurshudyan}
\affiliation{Department of Astronomy, School of Physical Sciences, University of Science and Technology of China, Hefei, Anhui 230026, China}
\affiliation{CAS Key Laboratory for Research in Galaxies and Cosmology, University of Science and Technology of China, Hefei, Anhui 230026, China}
\affiliation{School of Astronomy and Space Science, University of Science and Technology of China, Hefei, Anhui 230026, China}
\affiliation{Institut de Ciencies de lEspai (CSIC), Campus UAB, Carrer de Can Magrans, s/n 08193 Cerdanyola del Valles, Barcelona, Spain}
\affiliation{International Laboratory for Theoretical Cosmology, Tomsk State University of Control Systems and Radioelectronics, 634050 Tomsk, Russia}

\author{Emmanuel N. Saridakis}
\affiliation{Department of Physics, National Technical University of Athens, 
Zografou Campus GR 157 73, Athens, Greece}
\affiliation{National Observatory of Athens, Lofos Nymfon, 11852 Athens, 
Greece}
\affiliation{Department of Astronomy, School of Physical Sciences, University of Science and Technology of China, Hefei, Anhui 230026, China}

%\correspondingauthor{Yi-Fu Cai}
\email{yifucai@ustc.edu.cn}
%\correspondingauthor{Martiros Khurshudyan}
\email{khurshudyan@ustc.edu.cn}
%\correspondingauthor{Emmanuel N. Saridakis}
\email{msaridak@phys.uoa.gr}

\begin{abstract}
We apply Gaussian processes and Hubble function data in $f(T)$ cosmology, to reconstruct for the first time the $f(T)$ form in a model-independent way. In particular, using $H(z)$ datasets coming from cosmic chronometers as well as from the radial BAO method, alongside the latest released local value $H_{0} = 73.52 \pm 1.62$ km/s/Mpc, we reconstruct $H(z)$ and its derivatives, resulting eventually in a reconstructed region for $f(T)$, without any assumption. Although the cosmological constant lies in the central part of the reconstructed region, the obtained mean curve follows a quadratic function. Inspired by this we propose a new $f(T)$ parametrization, i.e. $f(T) = -2\Lambda +\xi T^2$, with $\xi$ the sole free parameter that quantifies the deviation from $\Lambda$CDM cosmology. Additionally, we confront three viable one-parameter $f(T)$ models of the literature, which respectively are the power-law, the square-root exponential, and the exponential one, with the reconstructed $f(T)$ region, and then we extract significantly improved constraints for their model parameters, comparing to the constraints that arise from usual observational analysis. Finally, we argue that since we are using the direct Hubble measurements and the local value for $H_0$ in our analysis,  with the above reconstruction of $f(T)$, the $H_0$ tension can be efficiently alleviated.
\end{abstract}

\keywords{cosmological parameters, gaussian process, test of modified gravity}

\section{Introduction} \label{sec:intro}

Modified gravity is an effective approach to describe the early- and late-time 
universe acceleration \citep{Capozziello:2011et,Nojiri:2010wj}), except for an 
introduction of the inflaton and/or dark energy components 
\citep{Peebles:2002gy, Cai:2009zp,Shafieloo:2019as,Li:2019xll,Martiros:2018emn}. 
In particular, the interest of study upon modified gravity has been revoked in 
particular by the recently reported measurements of the $H_0$ tension that has 
failed to be addressed within the standard $\Lambda$CDM cosmology 
\citep{Wong:2019kwg, Freedman:2019jwv, Vagnozzi:2019ezj}. Among various 
constructions of modified gravity one can find the interesting class that is 
based on the torsional formulation (for a review see \citep{Cai:2015emx}). In 
particular, starting from the simplest torsional gravity, namely 
Teleparallel Equivalent of General Relativity (TEGR) \citep{ein28, Hayashi79, 
Pereira.book, Maluf:2013gaa}, one can construct modifications such as the $f(T)$ 
gravity \citep{Cai:2015emx, Bengochea:2008gz, Linder:2010py, Chen:2010va, 
Myrzakulov:2010tc, Bamba:2010wb, Dent:2011zz, Cai:2011tc,Capozziello:2011hj, 
Wu:2011kh, Wei:2011aa, Amoros:2013nxa, Otalora:2013dsa, Bamba:2013jqa, 
Li:2013xea, Ong:2013qja, Nashed:2014lva, Darabi:2014dla, Haro:2014wha, 
Hanafy:2014ica, Guo:2015qbt,Bamba:2016gbu, Malekjani:2016mtm, Farrugia:2016qqe, 
Qi:2017xzl, Bahamonde:2017wwk, Karpathopoulos:2017arc, Cai:2018rzd, Li:2018ixg, 
Krssak:2018ywd, Iosifidis:2018zwo, nunes:2018rcn}, the $f(T,T_G)$ gravity 
\citep{Kofinas:2014owa, Kofinas:2014daa}, scalar-torsion theories 
\citep{Geng:2011aj, Hohmann:2018rwf}, etc. Finally, one can proceed 
going beyond the teleparallel 
framework, and construct more complicated torsional theories.

One key question for any theory of modified gravity is how to determine a viable choice from the arbitrary functions that have been involved. Some general features may be determined by theoretical arguments such as the theoretical requirements for a ghost-free theory that possesses stable perturbations etc,  or the desire for the action to possess Noether symmetries; however, the main tool of constraining the possible forms of modification remains that of observational confrontation. The general recipe is to consider by hand a variety of specific forms inside some general class, apply them in a cosmological framework, predict the dynamical behaviours at both the background and perturbation levels, and then use observational data to constrain the involved parameters or exclude the examined form (a similar procedure can also be followed to confront with local/solar system data). For the case of torsional gravity, the cosmological confrontations have been performed in \citep{Wu:2010mn, Nesseris:2013jea, Capozziello:2015rda, Basilakos:2016xob, Nunes:2016qyp, Nunes:2016plz, Nunes:2018xbm, Basilakos:2018arq, Anagnostopoulos:2019miu}, and the solar system tests can be found in \citep{Iorio:2012cm, Iorio:2015rla, Farrugia:2016xcw}, while the latest limit from galaxy lensing has been presented in \citep{Chen:2019ftv}. Hence,  in the literature there exist at least three viable scenarios for $f(T)$ gravity \citep{Nesseris:2013jea, Basilakos:2018arq}.

Although the above procedure of observational constraints is very useful to offer crucial information on the possible modification forms, it is even more expecting if the data can reconstruct the involved modification functions in a model-independent way without inserting an initial guess. Such a procedure has been successfully developed in the early-time inflationary cosmology \citep{Lidsey:1995np, Copeland:1993jj, Herrera:2018mvo}; however, concerning the late-time cosmology the involved complications allow only for a partial application, such as in the cosmography framework \citep{Bamba:2012cp, Capozziello:2019cav} or back-scattering procedure \citep{Capozziello:2017uam}. Interestingly, a useful tool towards the above reconstruction is the 
analysis of Gaussian processes (GP) \citep{RasmussenBook, Holsclaw:2010sk, Seikel:2013fda}, which allow one to investigate features of the form of the involved unknown functions in a model-independent way, using only the given datasets. Such a procedure has been applied to dark energy models 
using various datasets in order to reconstruct the evolution of the Hubble function, of the dark energy equation-of-state (EoS) parameters, of the dimensionless comoving luminosity distance, of the dark interaction term, etc. \citep{Martiros:2018emn, Seikel:2012uu, Shafieloo:2012ht, Seikel:2013fda, Kim:2013bja, Nair:2013sna, Cai:2015zoa, Costa:2015lja, Zhang:2016tto, Wang:2017jdm, Gomez-Valent:2018hwc, Pinho:2018amp, Elizalde:2018dvw, Melia:2018tzi, Pinho:2018unz, Zhang:2018gjb, Yin:2018mvu, Rau:2019zos, Gomez-Valent:2019lny}.

In this article we develop the GP analysis for the case of $f(T)$ cosmology, in order to reconstruct the form of the $f(T)$ modification in a model-independent way, namely using as the only input the observational datasets of Hubble function measurements $H(z)$. Such a procedure becomes 
easy in the case of $f(T)$ cosmology, since the latter has the advantage that the torsion scalar is a simple function of $H$, namely $T=-H^2$, and thus eventually all cosmological equations can be expressed in terms of $H(z)$ and its derivative. Hence, reconstructing $H(z)$ and its derivative 
from the Hubble data through the GP analysis, leads to the reconstruction of the $f(T)$ form itself, without any assumption.

The plan of the article is as follows. In Section \ref{FTsection} we provide a brief review of the cosmological equations of $f(T)$ gravity. In Section \ref{GaussianProcess} we describe the basic ingredients of GP and then we apply it for Hubble function observational data reconstructing $H(z)$ and its derivative. Then in Section \ref{ReconstructingfT} we use this reconstruction in order to reconstruct the form of $f(T)$ in a model-independent way, and to extract constraint on various $f(T)$ models in the literature. Finally, in Section \ref{Conclusions} we summarize our results with a discussion.

\section{$f(T)$ gravity and cosmology}
\label{FTsection}

In this section we briefly review $f(T)$ gravity and cosmology. In torsional 
formulation one uses as dynamical variables the vierbeins fields, which form an 
orthonormal basis at a manifold point. In a coordinate basis they are related to 
the metric through $g_{\mu\nu}(x)=\eta_{AB} ~ e^A_\mu (x) ~ e^B_\nu (x)$, with 
Greek and Latin indices respectively used for the coordinate and tangent space. 
In the particular class of teleparallel gravity one introduces the  
Weitzenb\"{o}ck connection  
$ {W}^\lambda_{\nu\mu}\equiv e^\lambda_A\: \partial_\mu e^A_\nu$ \citep{Weitzenb23}, and thus the corresponding torsion tensor is
\begin{equation}
\label{torsten}
{T}^\lambda_{\:\mu\nu} \equiv {W}^\lambda_{\nu\mu} - {W}^\lambda_{\mu\nu} = e^\lambda_A \: 
(\partial_\mu e^A_\nu - \partial_\nu e^A_\mu ) ~,
\end{equation}
while the corresponding curvature tensor is zero.
We mention here that the above formulation is constructed in a specific 
cosmological gauge where the spin connection components are vanishing, which 
simplifies the analysis 
and allows us to impose the specific vierbein choice below. Leaving a non-zero 
spin connection is also possible, however this would involve more complicated 
vierbeins \citep{Krssak:2015oua, Hohmann:2018rwf}.

The torsion tensor incorporates all the information of the gravitational field, and the torsion scalar arises from its contraction as 
\begin{equation}
\label{torsiscal}
T\equiv\frac{1}{4} T^{\rho \mu \nu} T_{\rho \mu \nu} + \frac{1}{2}T^{\rho \mu \nu }T_{\nu \mu \rho } 
- T_{\rho \mu }^{\ \ \rho }T_{\ \ \ \nu }^{\nu\mu} ~.
\end{equation}
This forms the Lagrangian of teleparallel gravity (similarly to general relativity where the Lagrangian is the Ricci scalar), and since variation in terms of the vierbeins gives the same equations with general relativity,  the constructed theory was named teleparallel equivalent of general relativity (TEGR).

One can start from TEGR to proceed the torsional based modifications. The simplest extension is to generalize $T$ in the action to be $T+f(T)$ \citep{Cai:2015emx},
\begin{eqnarray}
\label{action0}
S = \frac{1}{16\pi G}\int d^4x e \left[T+f(T)+L_m\right],
\end{eqnarray}
where $e = \text{det}(e_{\mu}^A) = \sqrt{-g}$, $G$ is the gravitational constant, and where for completeness we have also included the matter Lagrangian $L_m$. Variation of the above action results to the field equations
\begin{eqnarray}
\label{equationsom}
&&e^{-1}\partial_{\mu}(ee_A^{\rho}S_{\rho}{}^{\mu\nu})[1+f_{T}] + 
e_A^{\rho}S_{\rho}{}^{\mu\nu}\partial_{\mu}({T})f_{TT} \ \ \ \ \  \ \ \ \  \ \
\ \ \nonumber\\
&& \ \ \ \ 
-[1+f_{T}]e_{A}^{\lambda}T^{\rho}{}_{\mu\lambda}S_{\rho}{}^{\nu\mu}+\frac{1}{4} e_ {A}^{\nu}[T+f({T})] \nonumber \\
&&
\ \ \ \ \,
= 4\pi Ge_{A}^{\rho}\overset {\mathbf{em}}T_{\rho}{}^{\nu} ~,
\end{eqnarray}
with $f_{T}\equiv\partial f/\partial T$, $f_{TT}\equiv\partial^{2} f/\partial T^{2}$, and where $\overset{\mathbf{em}}{T}_{\rho}{}^{\nu}$ denotes the total matter (namely dark and baryonic matter) energy-momentum tensor. Finally, 
$S_\rho^{\:\:\:\mu\nu}\equiv\frac{1}{2}\Big(K^{\mu\nu}_{\:\:\:\:\rho}
+\delta^\mu_\rho \:T^{\alpha\nu}_{\:\:\:\:\alpha}-\delta^\nu_\rho\: 
T^{\alpha\mu}_{\:\:\:\:\alpha}\Big)$ is the super-potential, with 
$K^{\mu\nu}_{\:\:\:\:\rho}\equiv-\frac{1}{2}\Big(T^{\mu\nu}_{\:\:
\:\:\rho} - T^{\nu\mu}_{\:\:\:\:\rho}-T_{\rho}^{\:\:\:\:\mu\nu}\Big)$ the contorsion tensor. 

In order to apply $f(T)$ gravity at a cosmological framework we impose the spatially flat Friedmann-Lema\^{i}tre-Robertson-Walker (FLRW) metric 
\begin{equation}
ds^2= dt^2-a^2(t)\,  \delta_{ij} dx^i dx^j ~,
\end{equation}
which corresponds to the vierbein form $e_{\mu}^A={\rm diag}(1,a,a,a)$, where $a(t)$ is the scale factor. Inserting this choice into the general field equations (\ref{equationsom}) we result to the Friedmann equations of $f(T)$ cosmology, namely
\begin{align}
\label{Fr11}
&H^2= \frac{8\pi G}{3}\rho_m - \frac{f}{6} + \frac{Tf_T}{3} ~, \\
\label{Fr22}
&\dot{H}=-\frac{4\pi G(\rho_m+P_m)}{1+f_{T}+2Tf_{TT}} ~,
\end{align}
with $H\equiv\dot{a}/a$ the Hubble function and where dots denoting derivatives with respect to $t$. Additionally, in the above equations $\rho_m$ and $P_m$ are respectively the energy density and pressure of the matter fluid. Note that the torsion scalar (\ref{torsiscal}) in FRW geometry becomes 
\begin{eqnarray}\label{TscalarFRW}
T=-6H^2 ~,
\end{eqnarray}
and this expression proves very useful for the purpose of the present work.

As a next step we  define an effective dark energy sector with energy density and pressure respectively given by
\begin{align}
\label{rhoDDE} 
&\rho_{DE} \equiv \frac{3}{8\pi G} \Big[ -\frac{f}{6}+\frac{Tf_T}{3} \Big] ~, \\
\label{pDE}
&P_{DE} \equiv \frac{1}{16\pi G} \Big[ \frac{f-f_{T} T +2T^2f_{TT}}{1+f_{T}+2Tf_{TT}} 
\Big] ~, 
\end{align}
and therefore its EoS parameter reads as
\begin{eqnarray}
\label{wefftotf}
 w\equiv\frac{P_{DE}}{\rho_{DE}} = 
-\frac{f/T-f_{T}+2Tf_{TT}}{\left[1+f_{T}+2Tf_{TT}\right]\left[f/
T-2f_{T} \right] } ~.
\end{eqnarray}
Hence, the first Friedmann equation (\ref{Fr11}) effectively acquires the standard form $H^2=\frac{8\pi G}{3}(\rho_m+\rho_{DE})$. Finally, the system of cosmological equations closes by considering the matter conservation
\begin{eqnarray}
\label{mattradevol}
 \dot{\rho}_m+3H(\rho_m+P_m)=0 ~,
\end{eqnarray}
which using (\ref{Fr11}), (\ref{Fr22}) implies additionally the conservation of the effective dark energy,
\begin{eqnarray}
\label{evoleqr}
\dot{\rho}_{DE}+3H(\rho_{DE}+P_{DE})=0 ~.
\end{eqnarray}

\section{Gaussian Process using $H(z)$ data}
\label{GaussianProcess}

In this section we first present the general steps of the GP approach, and then we apply it in the case where the inserted data come from Hubble function observations. 
  
\subsection{Gaussian Process}

The GP is a powerful tool allowing to reconstruct the behavior of a function (and its derivatives)  directly from given datasets \citep{Seikel:2012uu}. The basic ingredients of the GP techniques are the covariance function (kernel), and the feature that the parameters describing it can be estimated directly from observational data \citep{Seikel:2012uu}. Hence, one does not need to consider any specific parametrization for the involved unknown function of the model, since it can be reconstructed from observational data directly by using the cosmological equations. 

In GP one assumes that the observations of the dataset are sampled from a multi-variance Gaussian distribution. Moreover, the values of the function evaluated at different points are not independent, and the connection between neighbouring points is due to the covariance functions chosen in advance. The Gaussian distribution corresponds to a random variable characterized by a mean value and a covariance. Similar to Gaussian distributions, GPs should be understood as distributions over functions, determined by a mean function and a covariance matrix. Since the covariance function, for a given set of observations, can infer the relation between independent and dependent variables, using the covariance function the GP correlates the function at 
different points. 

There exists a number of possible choices for the covariance function, i.e., for the kernel, e.g. squared exponential, polynomial, spline, etc, used in various applications \citep{RasmussenBook}. Although there is a discussion on possible effects of the kernel choice on the results \citep{RasmussenBook, Seikel:2013fda} (in a similar way that there is a discussion of the possible effect of the covariance matrix choice on the usual observational fittings \citep{Eifler:2008gx, Morrison:2013tqa}), one commonly used choice, with good theoretical justification, is the squared exponential function \citep{RasmussenBook, Seikel:2013fda}
\begin{equation}
\label{eq:kernel1}
k(x,x^{\prime}) = \sigma^{2}_{f} e^{-\frac{(x-x^{\prime})^{2}}{2l^{2}} } ~,
\end{equation}
where $\sigma_{f}$ and $l$ are parameters known as hyper-parameters. These 
parameters represent the length scales in the GP. In particular,  $l$   
corresponds to the correlation length along which the successive $f(x)$ values 
are correlated, while to control the variation in $f(x)$ relative to the mean of 
the process we need the parameter $\sigma_{f}$. Therefore, the covariance 
between output variables will be written as a function of the input ones. We 
mention that the covariance is maximum for variables whose inputs are suitably 
close. Furthermore, as can be seen from (\ref{eq:kernel1}), the squared 
exponential function is infinitely differentiable, which proves to be a useful 
property in case of reconstructing higher-order derivatives. Additionally, the 
initial 
and effective approach adopted in GP to estimate the values of the 
hyperparameters, is based on the training by maximizing the likelihood, showing 
that the reconstructed function has the measured values at the data points.

Finally, we mention that the consideration of the squared exponential 
kernel is the most natural choice amongst various possibilities, given the   
assumption that the error 
distribution for the observational data is 
the Gaussian one \citep{RasmussenBook,Seikel:2013fda}.
{\footnote{Another choice can be the so-called Matern ($\nu = 9/2$) covariance 
function \citep{RasmussenBook,Seikel:2013fda}
\begin{align}
\label{eq:kernel2}
k_{M}(x,x^{\prime}) = & \sigma^{2}_{f} e^{-\frac{3|x-x^{\prime}|}{l} } \times \Big[ 1+ \frac{3 |x-
x^{\prime}|}{l} + \frac{27(x-x^{\prime})}{7l^{2}} \nonumber\\
& + \frac{18|x-x^{\prime}|^{3}}{7l^{3}} + \frac{27 (x-x^{\prime})^{4}}{35 l^{4}} \Big] ~, 
\nonumber 
\end{align}
with $\sigma_{f}$ and $l$ the hyper-parameters. However, its application leads 
to similar results with the squared exponential function (\ref{eq:kernel1}), 
namely coincidence within 1$\sigma$ \citep{Elizalde:2018dvw}. Indeed, repeating the analysis of the present work using this 
alternative kernel shows that the difference between the results is
within $3\%$. Therefore, taking into account this fact, in this article we 
proceed only with the squared exponential kernel choice, namely
(\ref{eq:kernel1}).}}

\subsection{$H(z)$ data}

In this work we will use GP techniques with Hubble data $H(z)$ ($1+z = a_0/a$, with $a_0=1$ the present scale factor). In particular, we use $30$-point samples of $H(z)$  arising from the differential evolution of cosmic chronometers, alongside $10$-point samples obtained from the radial BAO method, which allows to extend the data range up to $z = 2.4$, improving also the behavior at low redshifts. In Table \ref{tab:Table0} we present the above points as they appear in \citep{Zhang:2016tto}. We note that in principle the use of data points from different datasets should be avoided, however, as it was discussed in \citep{Seikel:2012uu}, the simultaneous use of the above two data sets gives the increased statistics that it is necessary for the correct application of the GP. Finally, concerning the value of the Hubble parameter at present $H_0$, we use the latest released local value at 2.4\% precision, namely $H_{0} = 73.52 \pm 1.62$ km/s/Mpc \citep{Riess:2016jrr}. 
We mention here that since we are using the direct Hubble measurements and the local value for $H_0$, in our analysis and hence in our $f(T)$ reconstruction, the $H_0$ tension is alleviated by construction.
\footnote{Accompanied with the $H_0$ as well as $\sigma_8$ tensions, another 
model-independent approach of alleviation can be achieved by virtue of the 
effective field theory approach as shown in \citep{Yan:2019gbw}.}

%We use the publicly available package GaPP (Gaussian Processes in Python) developed by Seikel et al. \citep{Seikel:2012uu}, and we apply it for the aforementioned $H(z)$ data. The result of this elaboration is the successful reconstruction of $H(z)$ and $H^{\prime}(z)$ (primes denote derivative with respect to the redshift $z$) in a model-independent way. In Figure \ref{fig:Fig0_2}, the numerically derived mean curves and their 1$\sigma$ errors are presented \textit{for $H_{0} = 73.52 \pm 1.62$ km/s/Mpc} 
%\footnote{Note that, although we have imposed the value $H_{0} = 73.52 \pm 1.62$ km/s/Mpc, the GP reconstruction itself provides an automatically tuned value around $73.226 \pm 1.492$ km/s/Mpc at $z=0$, which remains self consistent. 
%\textit{On the other hand, the GP reconstruction provides an automatically tuned value around $68.053 \pm 0.404$ km/s/Mpc at $z=0$, when $H_0 = 67.66 \pm 0.42$ km/s/Mpc reported by the Planck mission has been used. Indeed which also remains self consistent.}}. 
%These reconstructions shall be used to reconstruct the $f(T)$ forms in the next section.

\begin{figure}[h]
\centering
\includegraphics[width=1.05\linewidth]{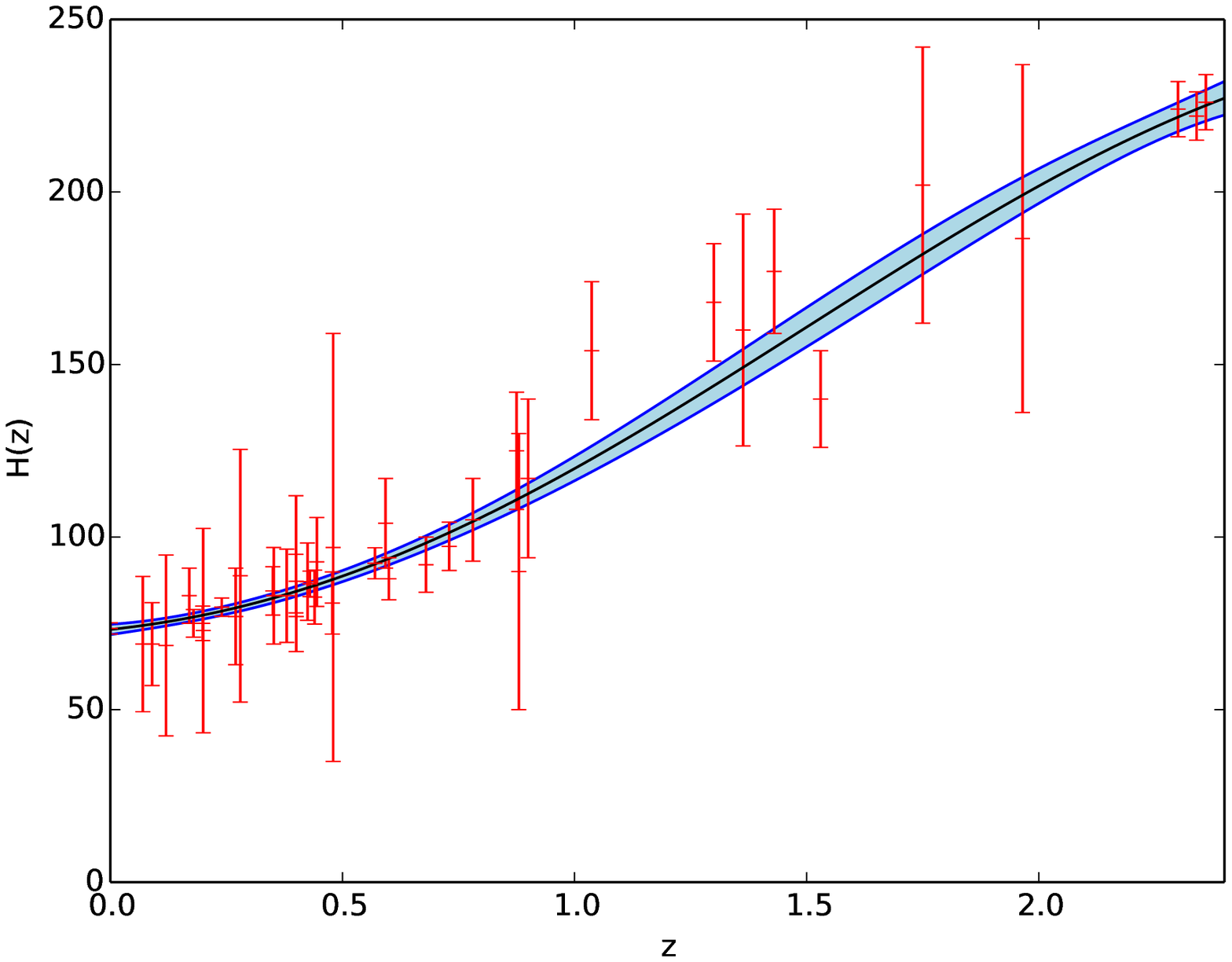}\\
\includegraphics[width=1.05\linewidth]{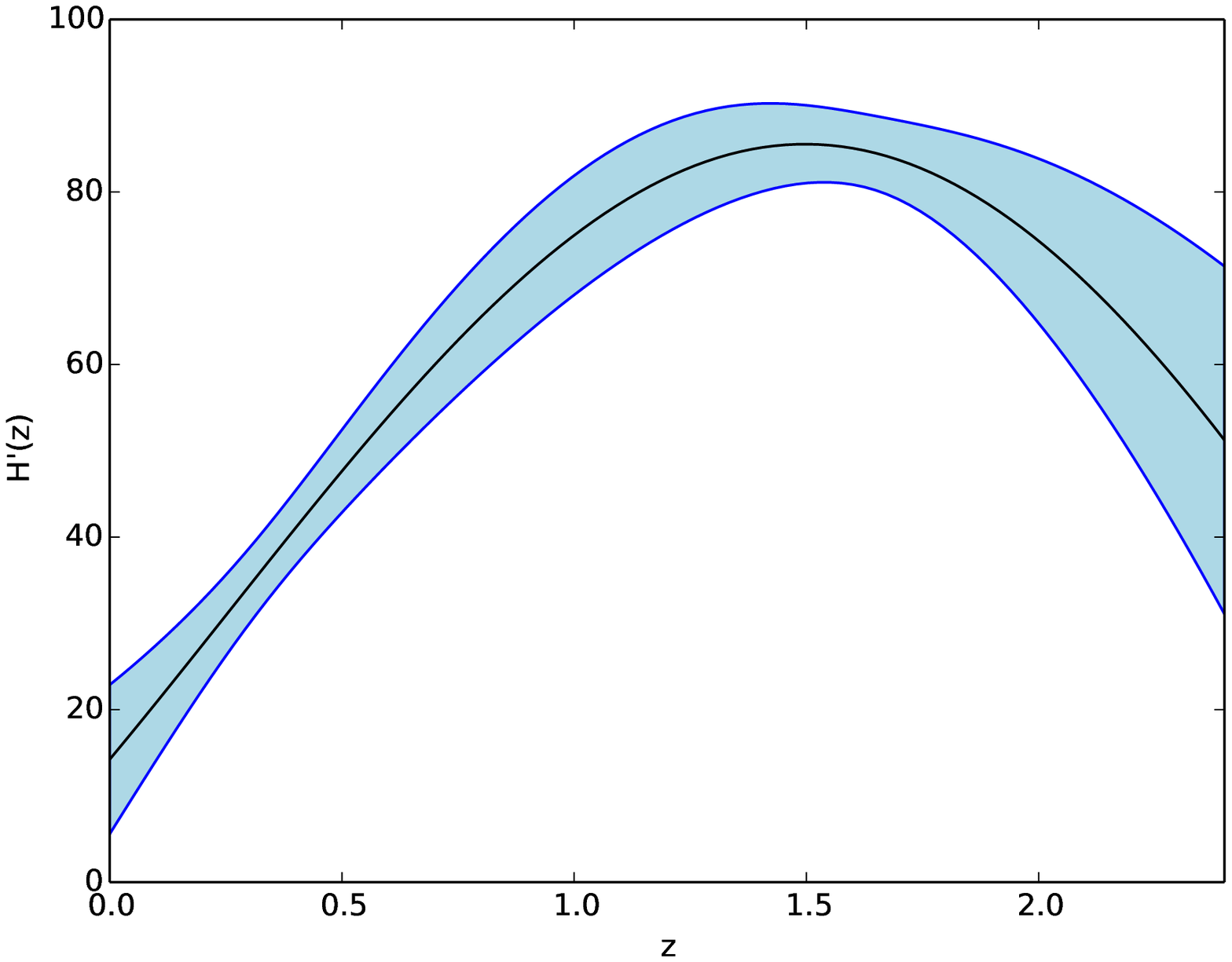}
\caption{{\it{GP reconstruction of $H(z)$ and $H^{\prime}(z)$ (primes denote derivative with respect to the redshift $z$), using the Hubble data arising from the differential evolution of cosmic chronometers ($30$-point sample) and from the radial BAO method ($10$-point sample) \citep{Zhang:2016tto}, presented in Table \ref{tab:Table0}, alongside the latest released local value  $H_{0} = 73.52 \pm 1.62$ km/s/Mpc \citep{Riess:2016jrr}, using the squared exponential kernel (\ref{eq:kernel1}). In each graph the black curve marks the mean reconstructed curve, while the light blue region marks the 1$\sigma$ errors coming from the data errors, as well as from the GP errors. We use units of km/s/Mpc.}}}
\label{fig:Fig0_2}
\end{figure}

We use the publicly available package GaPP (Gaussian Processes in Python) developed by Seikel et al. \citep{Seikel:2012uu}, and we apply it for the aforementioned $H(z)$ data. The result of this elaboration is the successful reconstruction of $H(z)$ and $H^{\prime}(z)$ (primes denote derivative with respect to the redshift $z$) in a model-independent way. In Figure \ref{fig:Fig0_2}, the numerically derived mean curves and their 1$\sigma$ errors are presented 
\footnote{Note that, although we have imposed the value $H_{0} = 73.52 \pm 1.62$ km/s/Mpc, the GP reconstruction itself provides an automatically tuned value around $73.226 \pm 1.492$ km/s/Mpc at $z=0$, which remains self consistent.}. 
These reconstructions shall be used to reconstruct the $f(T)$ forms in the next section.

\begin{table}[ht]
\centering
\begin{tabular}{ |  l   l   l  |  l   l  l  | p{2cm} |}
\hline
$z$ & $H(z)$ & $\sigma_{H}$ & $z$ & $H(z)$ & $\sigma_{H}$ \\
\hline
$0.070$ & $69$ & $19.6$ & $0.4783$ & $80.9$ & $9$ \\         
$0.090$ & $69$ & $12$ & $0.480$ & $97$ & $62$ \\
$0.120$ & $68.6$ & $26.2$ &  $0.593$ & $104$ & $13$  \\
$0.170$ & $83$ & $8$ & $0.680$ & $92$ & $8$  \\
$0.179$ & $75$ & $4$ &  $0.781$ & $105$ & $12$ \\
$0.199$ & $75$ & $5$ &  $0.875$ & $125$ & $17$ \\
$0.200$ & $72.9$ & $29.6$ &  $0.880$ & $90$ & $40$ \\
$0.270$ & $77$ & $14$ &  $0.900$ & $117$ & $23$ \\
$0.280$ & $88.8$ & $36.6$ &  $1.037$ & $154$ & $20$ \\
$0.352$ & $83$ & $14$ & $1.300$ & $168$ & $17$ \\
$0.3802$ & $83$ & $13.5$ &  $1.363$ & $160$ & $33.6$ \\
$0.400$ & $95$ & $17$ & $1.4307$ & $177$ & $18$ \\
$0.4004$ & $77$ & $10.2$ & $1.530$ & $140$ & $14$ \\
$0.4247$ & $87.1$ & $11.1$ & $1.750$ & $202$ & $40$ \\
$0.44497$ & $92.8$ & $12.9$ & $1.965$ & $186.5$ & $50.4$ \\

$$ & $$ & $$ & $$ & $$ & $$\\ 

$0.24$ & $79.69$ & $2.65$ & $0.60$ & $87.9$ & $6.1$ \\
$0.35$ & $84.4$ & $7$ &  $0.73$ & $97.3$ & $7.0$ \\
$0.43$ & $86.45$ & $3.68$ &  $2.30$ & $224$ & $8$ \\
$0.44$ & $82.6$ & $7.8$ &  $2.34$ & $222$ & $7$ \\
$0.57$ & $92.4$ & $4.5$ &  $2.36$ & $226$ & $8$ \\ 
\hline
\end{tabular}
\vspace{5mm}
\caption{The observational data for $H(z)$ and their uncertainty $\sigma_{H}$ in units of km/s/Mpc. In the upper panel we present the $30$ points deduced from the differential age method of cosmic chronometers, and in the lower panel we present $10$ samples obtained from the radial BAO method. The data is from \citep{Zhang:2016tto} (see references therein for each data point).}
\label{tab:Table0}
\end{table}

\section{Reconstructing $f(T)$ form from GP}
\label{ReconstructingfT}

In the previous section we applied the GP in the $H(z)$ data, and we resulted in the reconstruction of $H(z)$ and its derivatives, in a model-independent way, namely without assuming anything on the underlying gravitational theory or cosmological scenario. In this section we will assume that the universe is governed by $f(T)$ gravity, hence we will use the cosmological equations of Section \ref{FTsection}, and we will use the reconstructed  $H(z)$ and its derivatives to reconstruct the form of $f(T)$ without any other assumption. Up to our knowledge it is for the first time that this is performed, since up to now in the literature a specific ansatz for $f(T)$ was always imposed by hand and then the confrontation with the data led to constraining the involved parameters.

The basic feature of $f(T)$ gravity that makes the above reconstruction procedure easy is the fact that in FRW geometry for the torsion scalar we have relation (\ref{TscalarFRW}), namely $T=-6H^2$, i.e. it is a simple function of $H$. Thus, all the involved terms and functions of $f(T)$ cosmology can eventually be expressed in terms of $H(z)$ and its derivatives, which have been reconstructed through GP in the previous section. Hence, the final step of this analysis is the reconstruction of the form $f(T)$ itself. 

We start by expressing all cosmological equations of Section \ref{FTsection} in terms of the redshift $z$. For the time derivative of a function $h$ we have $\dot{h}=-h'H(1+z)$, where the prime denotes derivative with respect to $z$, while for $f_T$ we have  
\begin{equation}
f_{T} \equiv \frac{df(T)}{dT} = \frac{df/dz}{dT/dz}=\frac{f'}{T'} ~.
\end{equation}
The next step in the application of the GP  is to replace $f^{\prime}$ by
\begin{equation}\label{eq:fprime}
f^{\prime}(z) \approx \frac{f(z+\Delta z) - f(z)} {\Delta z} ~,
\end{equation}
for small $\Delta z$, which allows to relate the values of $f$ at $z_{i+1}$ and $z_{i}$. In particular, it is easy to see that from Equation (\ref{Fr11}) one acquires 
\begin{align}\label{eq:fz_rec0}
& f(z_{i+1}) - f(z_{i}) \\
& = 3(z_{i+1} \!-\! z_{i}) \frac{T^{\prime} (z_{i})}{T(z_{i})} \Big[ H^{2}(z_{i}) - 
\frac{8\pi G}{3}
 \rho_{m}(z_{i}) + \frac{f(z_{i})}{6}  \Big] ~, \nonumber
\end{align}
where $T=-6H^2$ and $T^{\prime} = - 12 H H^{\prime}$. Moreover, for the matter sector we adopt the EoS parameter for regular dust, which eventually gives
\begin{equation}
\rho_m = \frac{3}{8\pi G}H^{2}_{0} \Omega_{m0} (1+z)^{3} ~,
\end{equation}
due to (\ref{mattradevol}). Here $H_{0}$ and $\Omega_{m0}$ are the Hubble parameter and the dark matter density parameter ($\Omega_m = \frac{8\pi G \rho_m}{3H^2}$) at $z=0$. Inserting these into (\ref{eq:fz_rec0}) we finally obtain 
\begin{align}\label{eq:fz_rec}
& f(z_{i+1}) - f(z_{i}) \\
& = 6 (z_{i+1} - z_{i}) \frac{H^{\prime} (z_{i})}{H(z_{i})} \Big[ H^{2}(z_{i}) - H^{2}_{0} 
\Omega_{m0} (1+z_i)^{3} + \frac{f(z_{i})}{6} \Big] ~. \nonumber
\end{align}

Equation (\ref{eq:fz_rec}) allows us to reconstruct $f(z)$, as long as $H(z)$ and $H^{\prime}(z)$ are known. However, in the previous sections we were able to reconstruct $H(z)$ and $H^{\prime}(z)$ from the Hubble data. Thus, the reconstruction of  $f(z)$ is straightforward. Finally, since both $f(z)$ and $T(z)=-6 H^2(z)$ are reconstructed, we can easily reconstruct the form of $f(T)$ in a model-independent way. 

%In Figure \ref{fig:Fig0_3} we present the reconstructed $f(T)$ resulting from our analysis, \textit{when $H_0 = 67.66 \pm 0.42$ km/s/Mpc, which is the first main result of the present work.} Let us now try to extract information on the possible forms of $f(T)$. 

\begin{figure}[ht]
\centering
\includegraphics[width=1.1\linewidth]{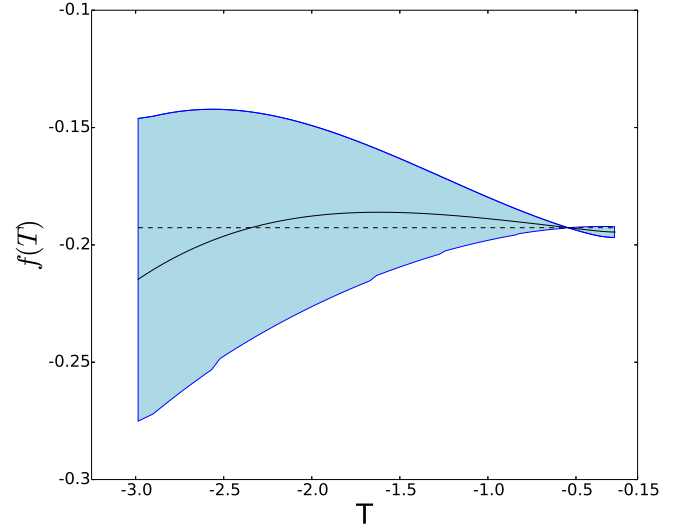} 
\caption{{\it{GP reconstruction of the $f(T)$ form in a model-independent way, using the reconstructed from Hubble data forms of $H(z)$ and $H^{\prime}(z)$ of Figure \ref{fig:Fig0_2} %with $H_{0} = 67.66 \pm 0.42$km/s/Mpc \citep{Aghanim:2018eyx} 
and the squared exponential kernel (\ref{eq:kernel1}), imposing 
$\Omega_{m0}=0.302$. The black curve marks the mean reconstructed curve, while 
the light blue region marks the 1$\sigma$ errors coming from  the GP errors. 
Moreover, the dotted line marks the cosmological constant scenario 
$f(T)=-2\Lambda=-6H_0^2(1-\Omega_{m0})$. Both $T$ and $f(T)$ are measured in 
units of $H^2$, i.e. $ (\text{km/s/Mpc})^2$, and we present them divided by 
$10^5$.
}}}
\label{fig:Fig0_3}
\end{figure}

In Figure \ref{fig:Fig0_3} we present the reconstructed $f(T)$ resulting from our analysis, which is the main result of the present work. Let us now try to extract information on the possible forms of $f(T)$. 

The first and clear result is that the $f(T)=-2\Lambda=const.$ form, namely the cosmological constant, lies in the central part of the reconstructed region, and in particular with the value $f(T)=-6H_0^2(1-\Omega_{m0})$ (the dotted line of Figure \ref{fig:Fig0_3}) which is exactly the cosmological constant one ($f(T)=-19267$ $ (\text{km/s/Mpc})^2$) as expected.

Nevertheless, from the reconstructed region of Figure \ref{fig:Fig0_3} we can obtain additional information, namely what is the form of the mean reconstructed curve (the black curve of Figure \ref{fig:Fig0_3}).
In particular, as we can see the mean curve is not a constant, but its best fit (with accuracy $R^2\approx0.94$) follows a quadratic function of the form $f(T) = -2\Lambda + \alpha T + \xi T^2$, with $-2\Lambda=-6H_0^2(1-\Omega_{m0})$ (i.e. $\Lambda$ is not a free parameter) and the two parameters $\alpha \approx -0.026 \pm 0.00088$, $\xi \approx (-9.68 \pm 0.28) \times 10^{-8}$ in units of km/s/Mpc. Since the linear term can be removed from $f(T)$ and be absorbed by the standard linear term that exists already in (\ref{action0}), the above form remains with one free parameter namely $\xi$, as it is the case in all viable models (that is why we did not allow for a second free parameter, namely $\gamma T^3$, although in this case the fitting is significantly improved at $R^2 \approx 0.999$). Hence, in summary, the mean curve of the reconstructed procedure follows the quadratic form 
\begin{equation}
\label{bestfit}
f(T) \approx -2 \Lambda+ \xi T^2 ~,
\end{equation}
with $\xi$ the sole free parameter. Note that, if we use a dimensionless free 
parameter then we may re-write (\ref{bestfit}) as $f(T) \approx-2\Lambda+ \beta 
T^2/T_0^2$, with $T_0=-6H_0^2$, and hence with the dimensionless parameter 
$\beta = 36 H_0^4 \xi$.

However, besides the mean curve of the reconstruction, in principle any curve inside the shaded region of Figure \ref{fig:Fig0_3} is allowed to be the true $f(T)$ form. Hence, let us confront three viable one-parameter $f(T)$ models of the literature \citep{Nesseris:2013jea, Basilakos:2018arq}, with our reconstructed region. In particular, they are  the power-law ($f_{1}$CDM) model 
\begin{equation}
f(T)=\alpha (-T)^{b} ~,
\label{powermod}
\end{equation} 
with $\alpha=(6H_0^2)^{1-b}\frac{1-\Omega_{m0}}{2b+1}$, the square-root exponential ($f_{2}$CDM) model  
\begin{eqnarray}
f(T)=\alpha T_{0}(1-e^{-p\sqrt{T/T_{0}}}) ~,
\label{Lindermod}
\end{eqnarray}
with $\alpha=\frac{1-\Omega_{m0}}{1-(1+p)e^{-p}}$ and $T_0=-6H_0^2$, and the exponential ($f_{3}$CDM) model
\begin{eqnarray}
f(T)=\alpha T_{0}(1-e^{-pT/T_{0}}) ~,
\label{f3cdmmodel}
\end{eqnarray}
with $\alpha=\frac{1-\Omega_{m0}}{1-(1+2p)e^{-p}}$. These models coincide with $\Lambda$CDM for $b=0$ (model (\ref{powermod})), and for $b=1/p \rightarrow 0^{+}$ (models (\ref{Lindermod}) and (\ref{f3cdmmodel})).

\begin{figure}[ht]
\centering
\includegraphics[width=1.1\linewidth]{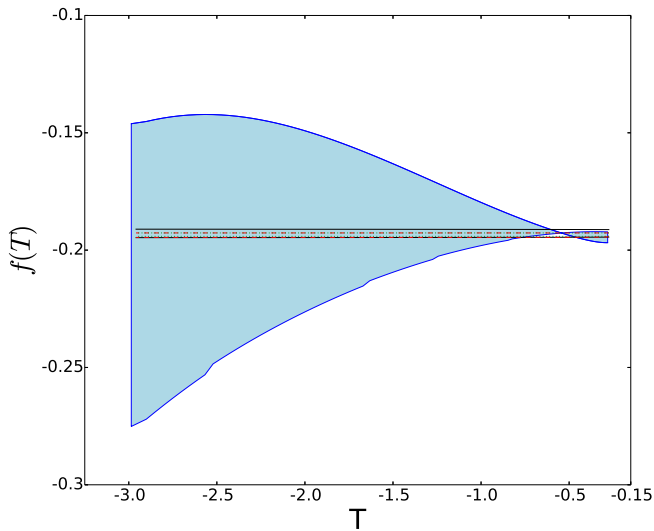}
\caption{{\it{GP reconstruction of the $f(T)$ form in a model-independent way, using the reconstructed from Hubble data forms of $H(z)$ and $H^{\prime}(z)$ of Figure \ref{fig:Fig0_2} %with $H_{0} = 67.66 \pm 0.42$km/s/Mpc \citep{Aghanim:2018eyx} 
and the squared exponential kernel (\ref{eq:kernel1}), imposing 
$\Omega_{m0}=0.302$. Additionally, we have added the predictions of three viable 
$f(T)$ models of the literature, for their edge parameter choices in order to 
still lie inside the reconstructed region, namely $b=-0.0005$ and $b=0.0004$ 
(black-solid curves) for the power-law model  (\ref{powermod}),  $b=1/p=0$ and 
$b=1/p=0.15$ (red-dashed curves) for the square-root exponential model  
(\ref{Lindermod}), and  $b=1/p=0$ and $b=1/p=0.13$ (green-dotted curves) for the 
 exponential model (\ref{f3cdmmodel}). Both $T$ and $f(T)$ are measured in units 
of $H^2$, i.e. $(\text{km/s/Mpc})^2$, and we present them divided by $10^5$.
}}}
 \label{fig:fTforms}
\end{figure}

\begin{figure}[!]
\centering
\includegraphics[width=1.0\linewidth]{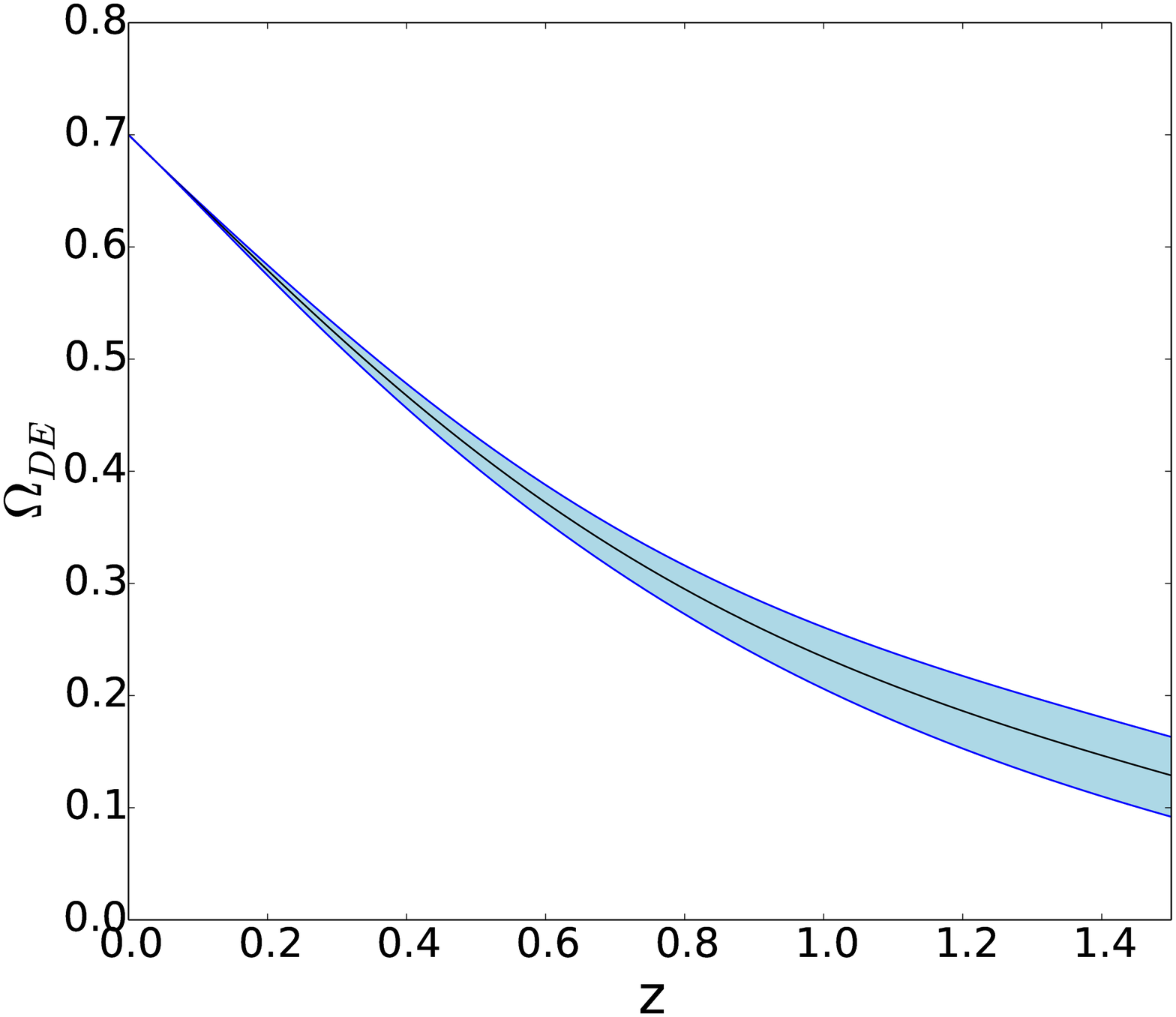}\\
\includegraphics[width=1.0\linewidth]{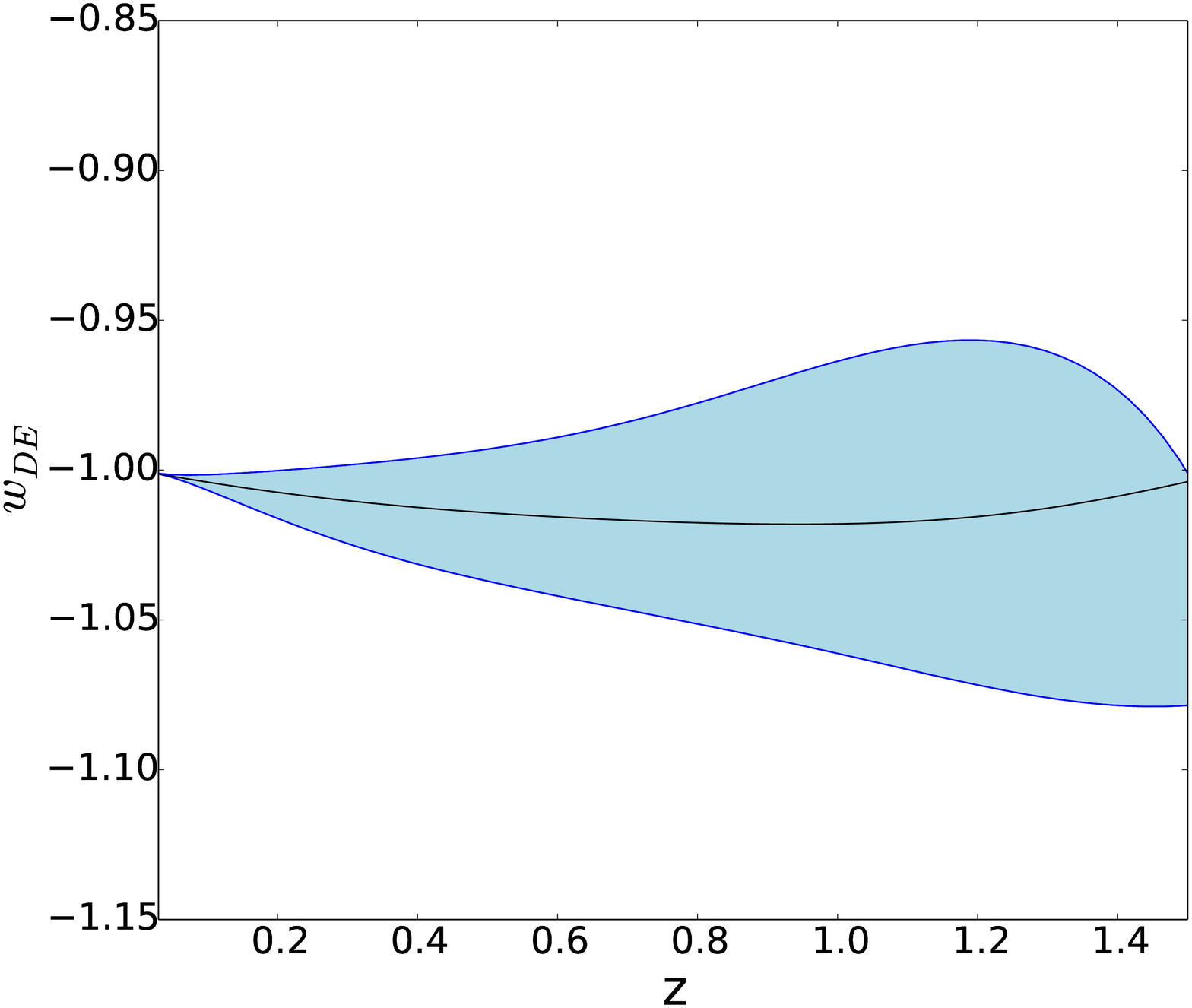}
\caption{{\it{The reconstructed forms of the dark energy density parameter $\Omega_{DE}$ from (\ref{rhoDDE}) (upper graph), as well as of the dark energy EoS parameter $w_{DE}$ from (\ref{wefftotf}) (lower graph), as they arise using the obtained reconstructions of $H(z)$, $H'(z)$ and $f(T)$. %when $H_{0} = 67.66 \pm 0.42$km/s/Mpc \citep{Aghanim:2018eyx}. 
In each graph the black curve marks the mean reconstructed curve, while the light blue region marks the 1$\sigma$ errors coming from the data errors, as well as from the GP errors.
}}}
 \label{fig:Fig0_4}
\end{figure}

As shown in Figure \ref{fig:fTforms}, if we expect the above $f(T)$ forms to lie inside the reconstructed region we obtain the following constraints on their sole parameter: $-0.0005<b<0.0004$ for the power-law ($f_{1}$CDM) model, $0<b=1/p<0.15$ for the square-root exponential ($f_{2}$CDM) model, and $0<b=1/p<0.13$ for the exponential ($f_{3}$CDM) model. Interestingly enough, these constraints are improved comparing to the constraints coming from usual observational analysis using Supernovae type Ia, quasi-stellar objects, Cosmic Microwave Background shift parameter, direct Hubble constant 
measurements with cosmic chronometers, and redshift space distortion ($f\sigma_8$) measurements, which give $-0.047<b<0.011$ for the power law, $-0.035<b=1/p<0.129$ for the square-root exponential, and $-0.011<b=1/p<0.111$ for the exponential models \citep{Basilakos:2018arq, Anagnostopoulos:2019miu}. Note the significant improvement by two orders of magnitude in the case of the power-law model. This is one of the main results of the present work, and shows the capabilities of the reconstruction procedure using GP.

Finally, concerning the quadratic form of $f(T)$ in Equation (\ref{bestfit}), we 
 deduce that in order to lie inside the reconstructed region its free parameter 
$\xi$ should be constrained as $-8.0 \times 10^{-8} < \xi < 4.5 \times 10^{-8}$ 
(in units of (km/s/Mpc)$^{-4}$). Equivalently, using the dimensional parameter 
$\beta = 36 H_0^4 \xi$ discussed above, we derive that $-59 < \beta < 33$. 
Hence, in the present work we propose the new $f(T)$ parametrization 
(\ref{bestfit}), since it can also be efficient in describing the reconstructed 
region that was obtained from the Hubble data through the GP analysis. The 
efficiency of the quadratic form was actually expected, since at late times 
where $H$ and thus $T$ are small, every function can be expanded in 
$T$-series 
\citep{Iorio:2012cm,Nashed:2015pda,Farrugia:2016xcw,
Bahamonde:2019zea,Chen:2019ftv}.

For completeness, we close this section by presenting in Figure \ref{fig:Fig0_4} the model-independent reconstructed forms of the dark energy density parameter $\Omega_{DE}={8\pi G \rho_{DE}}/{3H^2}$ from (\ref{rhoDDE}), as well as of the dark energy EoS parameter $w_{DE}$ from (\ref{wefftotf}), as they arise using the obtained reconstructions of $H(z)$, $H'(z)$ and 
$f(T)$. %when $H_{0} = 67.66 \pm 0.42$km/s/Mpc. 

\section{Conclusions}
\label{Conclusions}

In this work we have applied the GP analysis and Hubble function data in $f(T)$ cosmology, to reconstruct for the first time the $f(T)$ form in a model-independent way. In particular, up to now in the literature of $f(T)$ gravity, as well as in the majority of gravitational modifications, physicists were assuming specific ansatz for the involved unknown function, and were using observational data in order to constrain the model parameters. However, the use of the GP analysis allows to investigate features of the form of the involved unknown functions in a model-independent way without any assumption, using only the given observational datasets.

We applied the GP analysis for Hubble function measurements, namely for $H(z)$ datasets coming from cosmic chronometers  as well as from the radial BAO method, alongside the latest released local value  $H_{0} = 73.52 \pm 1.62$ km/s/Mpc at 2.4\% precision. Application of the procedure led to the reconstruction of $H(z)$ and its derivative without any assumption. On the other hand, $f(T)$ cosmology has the advantage that the torsion scalar is a simple function of $H$, namely $T=-H^2$, and thus eventually all cosmological equations can be expressed in terms of $H(z)$,$H'(z)$. Hence, having reconstructed  $H(z)$ and $H'(z)$ allowed us to additionally reconstruct  the $f(T)$ form itself, without any assumption. Up to our knowledge this is the first time where a general and model-independent reconstruction for the $f(T)$ gravity is obtained. Additionally, we mention that since we are using the direct Hubble measurements and the local value for $H_0$, in our analysis and hence in our $f(T)$ reconstruction, the $H_0$ tension can be alleviated by construction.

A first result of our analysis %, when $H_0 = 67.66 \pm 0.42$ km/s/Mpc, 
is that the cosmological constant lies in the central part of the reconstructed region, as expected. However, the  mean curve of the reconstructed region is not a constant, but its best fit  follows a quadratic function. Hence, inspired by this, in this work we proposed a new one-parameter $f(T)$ parametrization, namely $f(T) = -2\Lambda + \xi T^2$, with $-2\Lambda = -6H_0^2 (1-\Omega_{m0})$ and $\xi$ the sole free parameter that quantifies the deviation from $\Lambda$CDM  cosmology. Moreover,  fitting this form into the reconstructed region we extracted the constraints on the free parameter as $-8.0 \times 10^{-8} < \xi < 4.5 \times 10^{-8}$ (km/s/Mpc)$^{-4}$.

Additionally, we have confronted three viable one-parameter models of $f(T)$ with the reconstructed $f(T)$ region, which are the power-law, square-root exponential, and exponential one, respectively. As shown in the main text, we obtained improved constraints for their free parameters, comparing to the bounds that arise from traditional observational analyses, and especially for the case of the power-law model the improvement was more than two-orders of magnitude.

In summary, using GPs and Hubble data we obtained a model-independent reconstruction for the $f(T)$ form, and fitting its mean curve we proposed a new one-parameter $f(T)$ parametrization, namely the quadratic one. Finally, confronting three viable $f(T)$ models of the literature with our reconstructed region, we extracted improved constraints on their parameters. These features reveal the capabilities of the reconstruction procedure using GPs. Hence, it would be interesting to apply them in other theories of modified gravity too.

\acknowledgments
\section*{Acknowledgments}
We thank Canmin Deng, Rafael Nunes, Xin Ren, Yuting Wang and Gongbo Zhao for valuable comments.
We also thank the anonymous referee for constructive comments.
YFC is supported in part by the National Youth Thousand Talents Program of China, by NSFC (Nos. 11722327, 11653002, 11961131007, 11421303), by CAST (2016QNRC001), and by the Fundamental Research Funds for Central Universities.
MK is supported in part by a CAS President's International Fellowship Initiative Grant (No. 2018PM0054) and by NSFC (No. 11847226).
ENS is supported partly by USTC fellowship for international visiting professors. 
This article is partially based upon work from COST Action ``Cosmology and Astrophysics Network for Theoretical Advances and Training Actions'', supported by COST (European Cooperation in Science and Technology). 
All numerics were operated on the computer clusters {\it Linda \& Judy} in the particle cosmology group at USTC.

\end{document}